# Random lasing in an organic light-emitting crystal and its interplay with vertical cavity feedback

Andrea Camposeo[1]*, Marco Polo[1], Pompilio Del Carro[1], Leonardo Silvestri[2], Silvia Tavazzi[3], Dario Pisignano[1,4,**]

[1] National Nanotechnology Laboratory of Istituto Nanoscienze-CNR, via Arnesano I-73100, Lecce, Italy

[2] School of Electrical Engineering and Telecommunications, University of New South Wales, Sydney, NSW 2052, Australia

[3] Dipartimento Scienza dei Materiali, Università di Milano Bicocca, Via Cozzi 53, I-20125 Milano Italy

[4] Dipartimento di Matematica e Fisica "Ennio De Giorgi", Università del Salento, via Arnesano I-73100, Lecce, Italy

## Abstract

The simultaneous vertical-cavity and random lasing emission properties of a blue-emitting molecular crystal are investigated. The 1,1,4,4-tetraphenyl-1,3-butadiene samples, grown by physical vapour transport, feature room-temperature stimulated emission peaked at about 430 nm. Fabry-Pérot and random resonances are primed by the interfaces of the crystal with external media and by defect scatterers, respectively. The analysis of the resulting lasing spectra evidences the existence of narrow peaks due to both the *built-in* vertical Fabry-Pérot cavity and random lasing in a novel, surface-emitting configuration and threshold around 500 μJ cm$^{-2}$. The anti-correlation between different modes is also highlighted, due to competition for gain. Molecular crystals with optical gain candidate as promising photonic media inherently supporting multiple lasing mechanisms.

* Corresponding author. Tel: +390832298147. Fax: +390832298146. *E-mail address*: andrea.camposeo@nano.cnr.it

** *E-mail address*: dario.pisignano@unisalento.it

## 1. Introduction

Crystalline organic materials have the potential to enhance the characteristics of photonic devices thanks to their improved electronic and optical properties [1-3] and exciton-photon coupling phenomena. Recent examples are the observation of room temperature strong exciton-photon coupling [4], polariton lasing [5] and superradiance from large coherence volumes [6] in organic semiconductor single crystals. Furthermore, the molecular packing in crystalline phases results in anisotropic optical and electronic properties, in turn determining anisotropic charge mobility [7] and light emission and propagation [8, 9]. These features are deserving increasing interest in the framework of solid-state lasers [10-12]. Following the observation of amplified spontaneous emission (ASE) in α-octithiophene [13], different molecular crystals have been found to exhibit optical gain in the visible range, with typical excitation fluences of the order of few hundreds of μJ cm$^{-2}$. Line-narrowing by stimulated resonance Raman scattering has been reported [14, 15], as well as multimode laser oscillations [16-18].

Most of optically-gaining molecular crystals show ASE and lasing predominantly from the crystal edges, due to self-waveguiding of the fluorescence which favours light amplification along the crystal slab. However, this feature, besides preventing the future development of vertical-cavity lasers [5, 19] and of hybrid architectures exploiting light-emitting diodes for pumping [20], may hinder fundamental photonic phenomena, including Fabry-Pérot feedback provided by the reflective crystal facets and its eventual interplay with so-called random lasing, which is due to scattered light amplified by stimulated emission [21-23]. Indeed, molecular crystals can be unique systems for investigating lasing effects in the weak-scattering regime [24], where both scattered [15, 25] and Fabry-Pérot feedback can co-exist because of the relatively lower refractive index compared to inorganics [26, 27]. In this Letter, we report on the lasing emission of a surface-emitting molecular crystal, where both Fabry-Pérot and scattered feedback are present, as well as the onset of random lasing. Indeed, the analysis of the lasing spectra by micro-photoluminescence (μ-PL) evidences the co-existence of both mechanisms in the surface-emitting configuration, with threshold around 500 $\mu J\ cm^{-2}$. The threshold of modes with random lasing component is higher than that for Fabry-Pérot modes by about 40%. The anti-correlation between different modes is also highlighted, due to competition for gain. Optically-gaining molecular crystals are therefore interesting active media supporting multiple lasing mechanisms without the need of complex fabrication methods as typically required for realizing micro- or external cavities.

## 2. Experimental Methods and Materials

### 2.1 Crystal growth

Single crystals of 1,1,4,4-tetraphenyl-1,3-butadiene (TPB) were grown by the physical vapor transport method [28] in a three-zone furnace by placing TPB powder in a glass crucible at the end of a 75 cm long and 25 mm wide glass tube. The powder was heated at 175°C, whereas the temperature at the middle and the opposite end was 170°C and 135°C, respectively. The nitrogen flux was 50 mL/min. Single crystal of the $\beta$ monoclinic phase were selected under the polarizing transmission optical microscope and by atomic force microscopy analysis, as discussed elsewhere [29].

The morphology of the crystalline TPB with thicknesses in the range 5-10 $\mu m$ (measured by a stylus profilometer, Dektak 6M) was investigated by atomic force microscopy (AFM), with a Multimode system equipped with a Nanoscope IIIa electronic controller (Veeco Instruments). The surface topography of samples was imaged in Tapping mode, by using Si cantilevers with a resonance frequency of 250 kHz. Fluorescence micrographs were collected by either a stereomicroscope (Leica) or an inverted microscope (Nikon). In both systems samples were excited by a UV Hg lamp and the emission was imaged by a charge coupled device (CCD) camera.

*2.2 Laser excitation and spectroscopy*

Time-resolved emission was investigated by optically pumping the crystals by a frequency doubled Ti:Sapphire femtosecond laser source ($\lambda$= 385 nm, pulse width about 200 fs). The samples were mounted in a vacuum chamber in order to minimize photo-oxidation damage, with the *c* unit-cell axis parallel to the polarization of the pumping laser. The collected emission was dispersed by a monochromator coupled with a streak camera equipped with a two dimensional CCD, providing both wavelength- and time-resolved data, with a time resolution of about 20 ps. Different measurements were performed at variable pumping intensity in order to rule-out any intensity-dependent effects potentially affecting the photoluminescence (PL) decay. ASE was characterized by pumping with the third harmonic of a *Q*-switched Nd:YAG microlaser ($\lambda$=355 nm, pulse width of 0.6 ns, repetition rate 100 Hz). The excitation beam was focused onto the sample by a spherical lens (*f*=200 mm) along the direction normal to the crystal exposed surface, and emission was collected along the same direction. The polarization of the pumping beam was aligned either along *b* or *c* unit-cell axes. For studying lasing, we used a μ-PL set-up which allows specific areas of TPB crystals to be pumped. The third harmonic beam of the Nd-YAG microlaser was coupled to an inverted microscope (IX61, Olympus) and focused onto the sample by a 10× objective (numerical aperture of 0.3, area of the focused beam about 0.01 mm$^2$). The light emitted by the TPB crystals was collected from the air/crystal interface by a lens (*f*=60 mm) and coupled to an optical fiber. The PL spectral properties were analysed by a monochromator (iHR320, Jobin Yvon with a 1200 grooves/mm grating), equipped with a CCD detector (Symphony, Jobin Yvon). These measurements were performed in air and at ambient temperature, without observing significant changes of emission within the typical time intervals of measurements. In order to calculate the Fourier Transform (FT) of the emission spectra, these were firstly represented in unit of the wavevector $k=2\pi/\lambda$ (μm$^{-1}$). The FT of these data then provides a spectrum as a function of the conjugated variable, *d*, which is associated to the optical path length, expressed in μm.

### 3. Results and Discussion

We study here the $\beta$-polymorph of TPB, which, due to the peculiar arrangement of molecules in the unit cell (Figure 1a-b), exhibits ASE mainly from the top and bottom exposed crystal surfaces [30]. Therefore these surfaces can constitute the end-mirrors of *built-in* Fabry-Pérot cavities, allowing nanopatterning and deposition processes to be avoided and gain and emission properties to be consequently fully preserved. The crystallographic and morphological details of $\beta$-TPB are reported and discussed in Ref. [29]. Figure 1c shows typical fluorescence micrographs of $\beta$-TPB crystals, evidencing large areas with uniform emission. To constitute a Fabry-Pérot cavity, the crystal facets imaged in Fig. 1c have to be smooth and defect-free, thus minimizing scattering losses. Alternatively, the scattering provided by possible, microscopic morphological inhomogeneities can be exploited to provide the necessary feedback for random lasing. In fact, AFM highlights the presence of large flat areas with typical size of tens of microns, with terraces corresponding to growth steps of uniform height (~1 nm corresponding to the size of the $a$ axis of the unit cell, Figure 1d), as well as other areas with surface defects (with typical height of tens of nm and surface density of 0.3 $\mu m^{-2}$, Fig. 1e), and sporadic cracks (Fig. 1f). Hence a light mean free path $l_s$>0.3 mm can be estimated taking into account the scattering cross section for a two-dimensional system, $\sigma$ (~$10^{-7}$-$10^{-6}$ cm in organic materials [31]). Moreover, in the transmission spectrum of our crystals (Fig. S1 of the Supporting Information) we do not find evidence of narrow peaks, as observed in other systems and attributed to localized, long-lived modes [32].

In Figure 2a, we show the PL spectra taken from the $bc$ exposed face upon pulsed UV excitation. The arrangement of the molecule transition dipole moments in the unit cell favours emission mainly out of the crystallographic $bc$ plane. The room temperature spontaneous emission is peaked at about 430 nm, corresponding to the (0-1) replica of a vibronic progression, the (0-0) one being visible as a weak shoulder in the high energy tail of the emission spectrum, due to re-absorption of the emitted light. The time evolution of the emission is shown in the inset of Fig. 2a.

The decay is almost mono-exponential with a characteristic lifetime $\tau_{PL}$=0.95 ns, which corresponds to a radiative decay rate of about 0.3 ns$^{-1}$ taking into account the emission efficiency [30]. Moreover, under intense pulsed optical pumping with a normally-incident laser beam, the crystals show ASE from its top surface and spectral line narrowing (Fig. 2a) with typical threshold excitation fluences of few hundreds of $\mu J$ cm$^{-2}$. Such line narrowing is strongly sensitive to the polarization of the incident pumping laser. In the investigated pumping fluence range (0.1-1.7 mJ cm$^{-2}$), the spectral line narrowing can be observed only for polarization of the pumping beam parallel to the axis of maximum absorption (*c*) for normally-incident light (Fig. 2a). In fact, these crystals feature a quite high stimulated emission cross-section ($10^{-15}$ cm$^2$) [30], allowing for the observation of spectral line narrowing even for short gain path length (of the order of the crystal thickness in our experiments). In our range of excitation fluences, the gain length, $l_g$, inversely proportional to the optical gain, is in the range 5-10 $\mu m$, i.e. much shorter than the light mean free path, $l_s$. The weak and long-range disorder in TPB is therefore expected to favour lasing, by trapping stimulated emission [33]. Moreover, the here investigated configuration is substantially different from typical ASE experiments, where samples are excited along a stripe and emission is collected from an edge, thus being assisted by self-waveguiding [13, 16-18].

By exciting selected areas of the TPB crystals with a focused laser beam, narrow peaks with linewidth<0.3 nm (limited by the instrumental resolution) can be observed at room temperature. These peaks are also observed in emission spectra averaged over many laser shots (about 100 in Fig. 2b). In order to better understand the underlying mechanisms, we performed a systematic investigation of the emission by means of a $\mu$-PL system, allowing specific crystal areas (P1, P2, P3 and P4 in Fig. 2c) to be selectively excited with 100 $\mu m$ spatial resolution (Fig. 2c). The results of this study are shown in Fig. 3, where the emission spectra collected by exciting the various regions are analysed as a function of the incident pumping fluence. In all the investigated areas, the number of emission peaks increases by increasing the excitation intensity (Fig. 3a-d). A well-defined threshold is found, for the occurrence of the narrow emission peaks, in the range 500-800 $\mu J$ cm$^{-2}$.

A linear increase of the emission intensity is observed below threshold, well matching the crystal spontaneous emission trend (open squares in Fig. 3e-h), whereas a change of the slope of the input-output characteristics is observed above threshold as typical of lasing emission.

Overall, two mechanisms can be at the origin of the observed laser peaks, the first being related to resonant feedback due to the Fabry-Pérot cavity formed by the top and bottom reflecting interfaces, whereas the second can be associated to feedback provided by scattering from the morphological features shown in Fig. 1e-f. To discriminate among the two mechanisms and assessing their possible interplay, we carry out a Fourier analysis of the emission spectra [24]. For Fabry-Pérot cavities, the FT of the emission spectrum is characterized by the presence of discrete peaks at positions of the conjugated variable, $d$, [34]: $d_n = mnL/\pi$, where $m$ is an integer and $n$ the refractive index of the active material ($n$=1.8±0.1), whereas a more complex series of peaks is typically observed in random lasers [24, 33]. Figure 3i-l shows the FT of the averaged emission spectra, evidencing the presence of a complex pattern in most of the investigated areas which suggests predominant random lasing with characteristic cavity lengths in the range 10-20 μm. These findings support the "random cavities" model as a mechanism explaining the origin of the observed random lasing, similarly to amorphous conjugated polymer films [33, 35]. Interestingly, in the region P2, the FT of the spectra collected at pumping fluences slightly above threshold features well-defined resonances spaced by about 5 μm, providing a cavity length roughly matching the local crystal thickness. In fact, the corresponding emission spectrum (Fig. S2a of the Supporting Information) is characterized by the presence of two emission peaks spaced by $\Delta\lambda = 5.8$ nm, as expected for a Fabry-Pérot cavity with a length given by the crystal thickness ($\Delta\lambda \cong \lambda_n^2 / 2nL$ =5-6 nm). These particular modes, characterized by the lowest measured excitation threshold fluence (500 μJ cm$^{-2}$), can be therefore associated to Fabry-Pérot resonances. At high excitation fluences (≥700 μJ cm$^{-2}$), both Fabry-Pérot and random modes are present (see also Fig. S3 of the Supporting Information). Such behaviour can be rationalized considering the threshold gain for both the considered lasing mechanisms. For Fabry-Pérot cavities, the threshold gain is given by $g_{th}^{FP} = L^{-1}\ln(1/\sqrt{R_1 R_2})$, where $L$ is the cavity length and $R_1$ and $R_2$ are the reflectances of the cavity mirrors (the crystal/air and crystal/substrate interfaces in our experiments). The threshold gain can be also estimated for random media of characteristic length $L_s$ (23 μm as derived from the FT spectra of Fig. 3j where resonances spaced about 13 μm are observed for excitation fluences> 700 μJ cm$^{-2}$), filled with weakly-scattering point centers [26]: $g_{th}^{RL} = L_s^{-1}\ln(2\pi L_s / \sigma)$. For the P2 region, we find $g_{th}^{RL} / g_{th}^{FP} \approx 1.4$, in accordance with the experimental findings, evidencing vertical-cavity lasing at low excitation intensity and the presence of both Fabry-Pérot and random lasing modes as the excitation fluence is increased. In other regions, such as P4 (where $L_s$=17 μm), we find $g_{th}^{RL} / g_{th}^{FP} \approx 1$. This is in agreement with the collected spectra, where both Fabry-Pérot and random modes are present even at the lowest excitation densities above threshold (Fig.s 3d, h and S2b).

Finally, in Fig. 4 the temporal evolution of averaged emission spectra are analysed. Here it is worth noting that the intensity fluctuations of the different emission peaks are in part correlated. For instance, by looking at the intensity fluctuations of the peaks located at $\lambda$=426.7 nm and $\lambda$=428.2 (Fig. 4b), an anti-correlation can be appreciated since a maximum in the temporal evolution of the intensity of one mode often corresponds to a minimum of the other one. This is quantified by calculating the statistical correlation coefficient between the modes, defined as: $C = \sum_{i=0}^{n-1}\left(I_i^{\lambda_1} - I_{avg}^{\lambda_1}\right)\left(I_i^{\lambda_2} - I_{avg}^{\lambda_2}\right) / \sqrt{\sum_{i=0}^{n-1}\left(I_i^{\lambda_1} - I_{avg}^{\lambda_1}\right)^2 \sum_{i=0}^{n-1}\left(I_i^{\lambda_2} - I_{avg}^{\lambda_2}\right)^2}$, where $I_i^{\lambda_{1,2}}$ are the intensity of a given lasing mode with wavelength $\lambda_{1,2}$ at time $t_i = t_0 + i \times \Delta t$ (in our experiments $t_0$=0 and $\Delta t$=0.1 s) and $I_{avg}^{\lambda_{1,2}}$ is the intensity averaged over the investigated time interval (5 s). For the modes studied in Fig. 4 we find $C = -0.1$, a negative value indicative of competition which may occur between spatially overlapping lasing modes, that have to compete for gain [36]. Hence this analysis provides an experimental evidence of the dynamics and coupling of lasing modes in the random gain medium.

## 4. Conclusion

In summary, we studied the emission features of a TPB molecular crystal upon pulsed optical pumping, revealing various aspects of multiple lasing mechanisms in organics. First, investigating spatially-selected sample regions allows to find evidence of multimode lasing. Second, the most of emission has the typical features of a random laser, associated to the scattering of the crystal morphological defects. Both built-in Fabry-Pérot modes and random lasing modes can simultaneously occur, with vertical Fabry-Pérot modes being dominant at low pumping fluences (500-700 $\mu$J cm$^{-2}$), whereas random lasing is dominant at higher fluences. Third, the various modes are anticorrelated due to the competition for available gain. These results pave the way for the realization of solid-state lasers exploiting various amplification routes, and particularly random lasing, based on organic crystals. These do not require the integration of micro- or external cavities, the feedback for multiple lasing being provided by the material structure.

## Acknowledgments

The research leading to these results has received funding from the European Research Council under the European Union's Seventh Framework Programme (FP/2007-2013)/ERC Grant Agreement n. 306357 (ERC Starting Grant "NANO-JETS"). We also acknowledge the support from the Apulia Regional Projects 'Networks of Public Research Laboratories', WAFITECH (09) and M. I. T. T. (13).

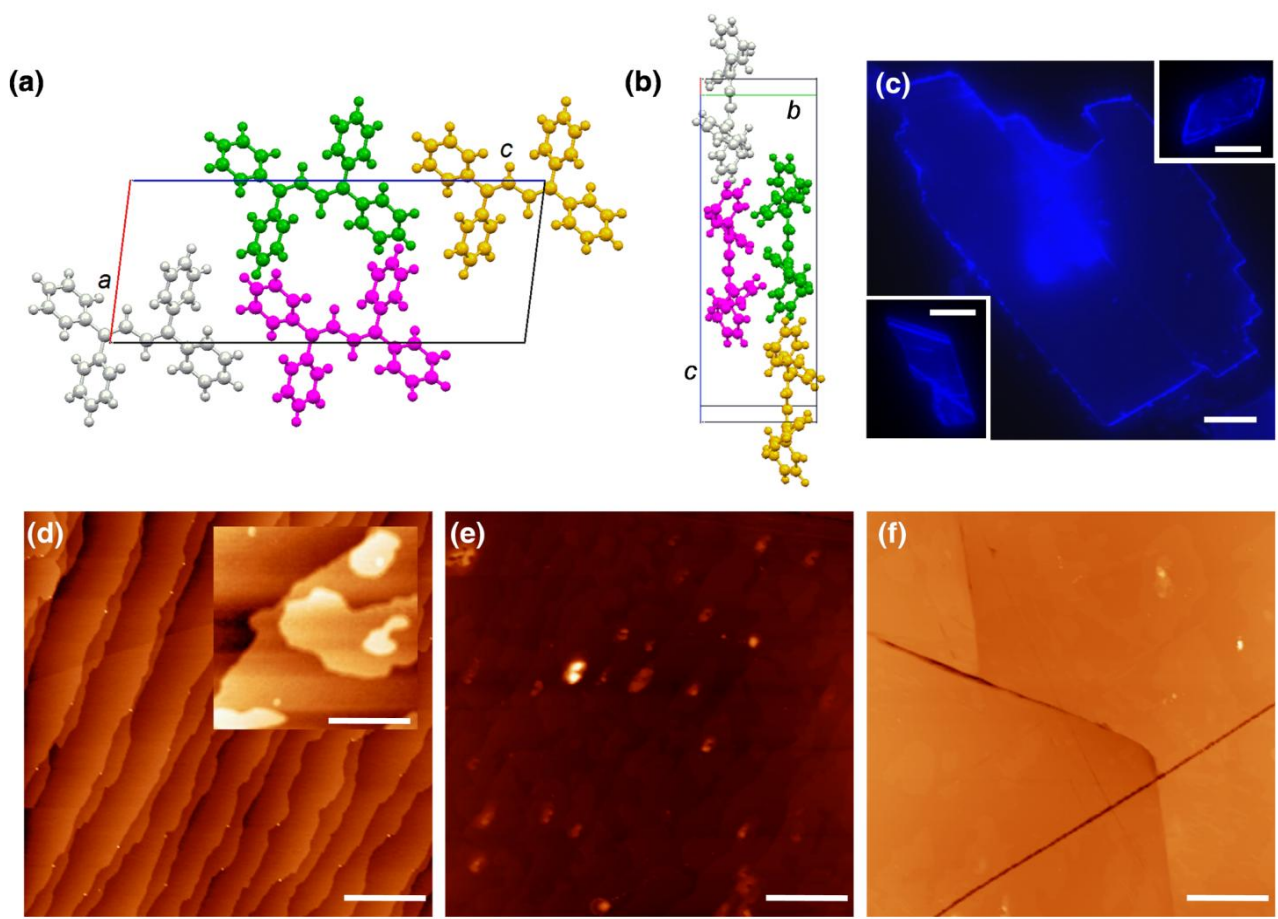

**Figure 1.** (a)-(b) Crystal structure and arrangement of TPB molecules in the unit cell, as viewed along the *b* axis (a) and along the normal (*a\**) to the *bc* plane (b), respectively. The unit cell contains four molecules (*Z* = 4) and its axes are *a* = 9.736 Å, *b* = 8.634 Å, *c* = 24.480 Å with angle *β* = (97.11)°. (c) Fluorescence micrographs of TPB crystals. Scale bars: 50 μm. The crystal exposed face corresponds to the *bc* crystal plane. (d)-(f) AFM surface topography maps of TPB crystals. The imaged crystal surface corresponds to the exposed surface (*bc* crystal plane). Scale bars: 2 μm. Inset in (d): image of growth steps with height of about 1 nm. Scale bar: 200 nm.

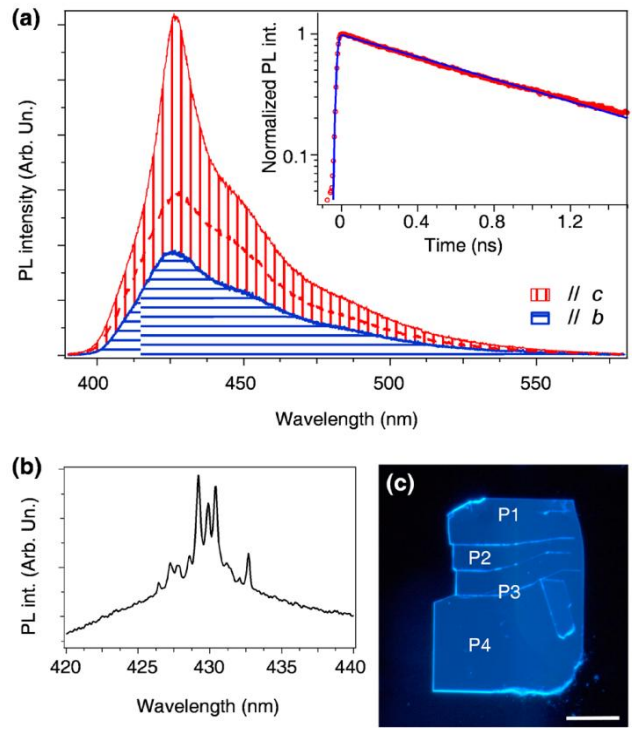

**Figure 2.** ASE dynamics. (a) ASE spectra acquired with incident pumping laser polarization parallel to the *c* axis (red vertically textured spectrum) and to the *b* axis (blue horizontally textured spectrum). The emission spectrum below ASE threshold with incident pumping laser polarization parallel to *c* axis is also shown (red dashed line). Inset: time evolution of the emission intensity at $\lambda$=430 nm. The continuous line is a fit to the data by an exponential decay curve, after convolution with a Gaussian instrumental response function with width of 20 ps. (b) Emission spectrum of a TPB crystal above the ASE threshold, evidencing the presence of sharp emission peaks. Excitation fluence = 1.1 mJ cm$^{-2}$. (c) Fluorescence micrograph of the crystals studied by μ-PL measurements. The investigated samples areas are highlighted as P1, P2, P3 and P4. Scale bar: 200 μm.

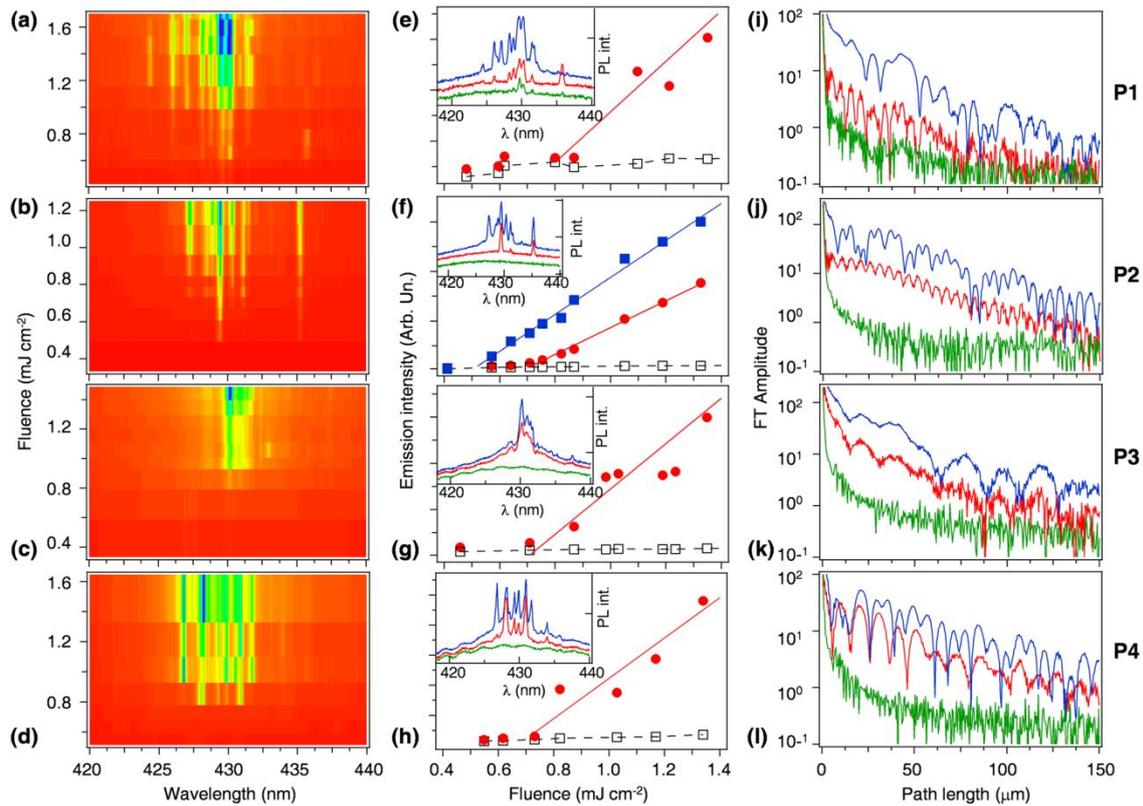

**Figure 3.** Lasing mode organization and structure. (a)-(d) Intensity maps showing the crystal emission spectra as a function of the excitation fluence. Data are collected as μ-PL, i.e. selectively exciting the areas P1, P2, P3, P4 (from top to bottom maps and plots, respectively). (e)-(h) Laser emission intensity *vs*. excitation fluence. Hollow squares show the intensity of the spontaneous emission in the different sample areas, that is linear in all the investigated range of excitation fluence. Data shown as solid symbols in (e), (f), (g) and (h) correspond to modes at $\lambda$=427.1 nm (e), 429.5 nm (full squares, f) and 430.2 nm (dots, f), 430.1 nm (g) and 428.0 nm (h), respectively. Insets: examples of emission spectra detected upon varying pumping fluences (from bottom to top spectra in the insets, the incident laser fluence is increased in the range 0.4-1.6 mJ cm$^{-2}$). The PL intensity in insets is shown in a logarithmic scale.(i)-(l) Amplitude of the emission spectra FT for various excitation fluences, acquired in the different sample areas. The original spectra are those the insets of (e)-(h), respectively. Incident fluences values from bottom to top: (i) 0.48, 0.80 and 1.60 mJ cm$^{-2}$; (j) 0.41, 0.64 and 1.05 mJ cm$^{-2}$; (k) 0.46, 0.98 and 1.44 mJ cm$^{-2}$; (l) 0.73, 0.82 and 1.20 mJ cm$^{-2}$.

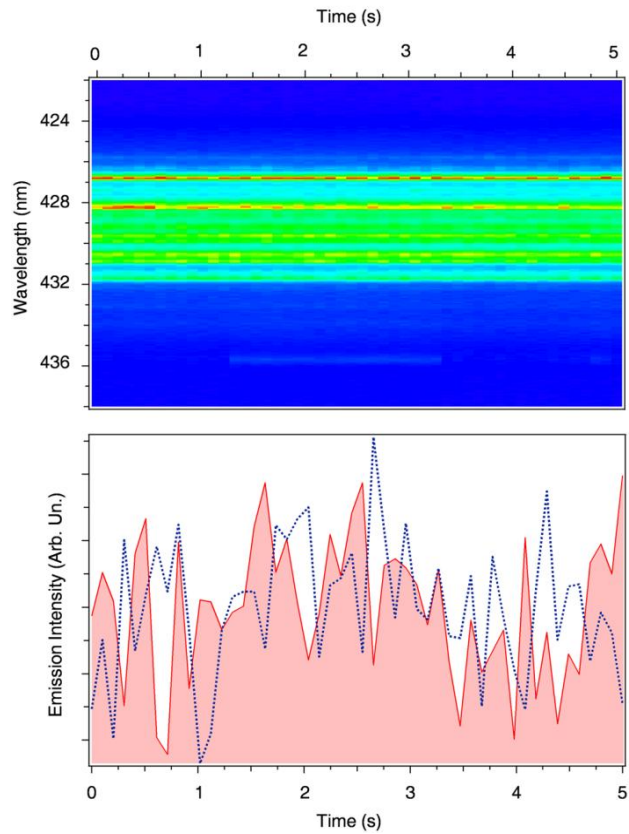

**Figure 4**. Competition between different modes in the Fabry-Pérot and random lasing cavity. (a) Temporal evolution of the lasing emission spectra above threshold (exemplary data from the P4 region, each spectrum is averaged over 10 laser shots). (b) Temporal evolution of the intensity of the lasing peaks at $\lambda$=426.7 nm (continuous line) and $\lambda$=428.2 nm (dotted line), respectively. Excitation fluence = 1.4 mJ cm$^{-2}$.